\documentstyle[preprint,aps]{revtex}

\begin{document}

\preprint{
\font\fortssbx=cmssbx10 scaled \magstep2
\hbox to \hsize{
\hskip.5in \raise.1in\hbox{\fortssbx University of Wisconsin - Madison}
\hfill$\vtop{\hbox{\bf MAD/PH/831}
                \hbox{\bf RAL-94-047}
                \hbox{\bf hep-ph/9405245}
                \hbox{May 1994}}$ }
}

\title{\vspace*{.75in}
Multilepton SUSY signals from $R$-parity violation\\ at the Tevatron}

\author{V. Barger$^a$, M.S.~Berger$^a$, P.~Ohmann$^a$ and R.J.N. Phillips$^b$}

\address{
$^a$Physics Department, University of Wisconsin, Madison, WI53706, USA\\
$^b$Rutherford Appleton Laboratory, Chilton, Didcot, Oxon OX11 0QX, UK}

\maketitle

\thispagestyle{empty}

\begin{abstract}
The expected trilepton signals from $p \bar p \to \chi^\pm_1\chi^0_2 \to
(\chi^0_1\ell^\pm\nu) (\chi^0_1\ell'^+\ell'^-) $ will be converted into
hadronically quiet multilepton signals, if the two final $\chi^0_1$ have
leptonic $R$-parity-violating (RPV) decays $\chi^0_1 \to \ell \ell' \nu$.
We make illustrative calculations of the acceptance for these spectacular
RPV signals, and point out that distinctive multilepton signals are possible
even when the $R$-conserving trilepton signals are
blocked by the ``spoiler mode" $\chi^0_2 \to h^0 \chi^0_1$.
Other channels such as $p\bar p\to \chi_1^\pm \chi_2^0 \to
(\chi_1^0\ell^\pm\nu) (\chi_1^0\nu\nu)$,
$p\bar p\to \chi_1^\pm \chi_1^0 \to (\chi_1^0\ell\nu)\chi_1^0$
and $p\bar p\to\chi_1^+\chi_1^-\to(\chi_1^0\ell^+\nu)(\chi_1^0\ell'^-\nu)$
can also give quiet multileptons from RPV.
We investigate these signals in the context of supersymmetric models
with radiative electroweak symmetry breaking, using examples in the
low-$\tan\beta$ $\lambda_t$ fixed-point region.
\end{abstract}

\newpage

There is intense interest in searching for signatures of
Supersymmetry (SUSY) at the Fermilab Tevatron $p\bar p$ collider, where
the highest present CM energy $\sqrt s = 1.8$--2~TeV is accessed.  An
important possibility is the pair production of charginos and
neutralinos\cite{brkp}, whose leptonic decay modes lead to many promising
signatures\cite{bkp,baer,nath,barbieri,lopez}.
Recently, much theoretical\cite{baer,lopez} and experimental\cite{cdf,d0}
attention has centered on trileptons from the production/decay
sequence
\begin{equation}
p \bar p \to \chi^\pm_1\chi^0_2 \to
(\chi^0_1 \ell^\pm\nu)(\chi^0_1 \ell'^+\ell'^-)\;.
\end{equation}
Here $\chi_i^\pm$ and $\chi_j^0$ are charginos and neutralinos ($i,j$ denote
mass ordering) and  $\chi^0_1$ is the lightest SUSY particle (LSP); see
Figs.~\ref{fig:feyns}(a) and \ref{fig:feyns}(b).  These trilepton events are
distinctively ``quiet" (little accompanying hadronic excitation); measurable
rates are predicted for interesting ranges of SUSY
parameters, but are lost in certain parameter regions e.g.\ where
the ``spoiler mode" $\chi^0_2 \to \chi^0_1 h$ is kinematically
accessible and suppresses all other $\chi^0_2$ decays ($h$ being the lightest
Higgs scalar), or where one of the leptons is constrained to be soft and
becomes undetectable\cite{baer,lopez}.

The popular scenario above assumes the LSP is stable and therefore
practically invisible, due to $R$-parity conservation (RPC).
The picture
changes dramatically with $R$-parity violation (RPV)\cite{dim}.  In particular,
if explicit RPV occurs through $L_iL_j\bar E_k$ terms in the superpotential,
the LSP will decay via  $\chi^0_1 \to \ell \ell' \nu$ to a neutrino plus two
charged leptons that may have different flavors (see Fig.~\ref{fig:feyns}(c))
thus converting the RPC trilepton signal into multileptons\cite{ross} with up
to seven charged leptons appearing in the final state.
Even when the spoiler mode is active, suppressing RPC trileptons,
a total of five charged leptons are still present in the decay of $\chi^\pm_1
\chi^0_2$ with RPV.   Also, the channels
$\chi_1^\pm \chi_2^0 \to (\chi_1^0\ell^\pm\nu)(\chi_1^0\nu\nu)$,
$\chi_1^\pm \chi_1^0 \to (\chi_1^0 \ell\nu)\chi_1^0 $
or $\chi_1^+\chi_1^-\to(\chi_1^0\ell^+\nu)(\chi_1^0\ell'^-\nu)$
give quiet signals with up to five or six leptons.
The actual multiplicity of observed leptons depends on the experimental
thresholds and angular acceptances; in the present paper we give
some sample calculations illustrating the high visibility of
these multilepton signals.

In the framework of the minimal supersymmetric standard model (MSSM)
with grand unification (GUTs),
the masses and couplings of the charginos $\chi^\pm_i$ and neutralinos
$\chi^0_j$ are determined by known quantities such as the gauge couplings
plus a number of parameters at the SUSY mass
scale: (a)~the gluino mass $m_{\tilde g}$; (b)~the Higgsino mass mixing
coefficient $\mu$;
(c)~the ratio of vacuum expectation values for the two Higgs doublets,
$v_2/v_1= \tan\beta$; (d)~the squark masses $m_{\tilde q}$ and
(e)~the CP-odd neutral Higgs mass $m_A$.
Unification constraints on gauge and Yukawa couplings at the
GUT scale lead to a greatly reduced parameter set at the SUSY
mass scale, through the renormalization group equations (RGE).
Such approaches
also explain electroweak symmetry breaking as a radiatively induced effect.
Recent analyses of the sparticle masses and couplings expected in supergravity
models can be found in Refs.\cite{guts,bbo,bbo801,lp,cpw,kane}. A particularly
attractive scenario is the occurence of an infrared
fixed point of the top-quark Yukawa
coupling\cite{pr}, which predicts the relation\cite{bbo}
\begin{equation}
m_t({\rm pole}) \simeq (200~{\rm GeV}) \sin\beta \;.
\end{equation}
In the following we adopt this scenario and
use the value $m_t=168$~GeV, for which the RGE solutions of
Ref.\cite{bbo801} were constructed, consistent with
$m_t = 174\pm10^{+13}_{-12}$~GeV from the CDF top-quark candidate
events\cite{top}. For this $m_t$ choice  $\tan\beta=1.5$.
Our analysis of the chargino and
neutralino signatures will be based on Ref.\cite{bbo801} where $m_0$ and
$m_{1/2}$ are input parameters at the GUT scale $M_G$ (along with trilinear
couplings $A=0$ at $M_G$), obtained with a naturalness condition
$|\mu|<0.5$~TeV on radiative symmetry breaking.
The allowed region in $(m_{1/2},m_0)$ parameter space is shown in Fig.~8 of
Ref.\cite{bbo801}.
The remaining parameters are determined by the RGE analysis within a twofold
ambiguity, corresponding to positive or negative $\mu$.

The fixed-point solution exhibits some simple characteristics,
which can be qualitatively understood from the tree-level relationship
\begin{equation}
{1\over2}M_Z^2 = {m_{H_1}^2 - m_{H_2}^2\tan^2\beta\over\tan^2\beta -1}-\mu^2\;.
\end{equation}
At the electroweak scale, $m_{H_2}^2\alt 0$ and hence $|\mu|$ must be large,
and in fact $|\mu|$ is found to be
substantially larger than $M_2$, the SU(2) gaugino mass.
Consequently the $\tilde W$-$\tilde H$ mass matrix is approximately diagonal
and the lightest chargino eigenstate $\chi_1^+$ is almost a pure $\tilde W^+$
state. Also, $\chi_1^0$ is nearly a pure U(1) gaugino $\tilde B$ while
$\chi_2^0$ is almost purely $\tilde W^0$. A direct result is that the
$W^+\chi_1^-\chi_2^0$ and the $Z\chi_1^+\chi_1^-$ couplings are
almost the maximal gauge couplings, and that the
$W^+\chi_1^-\chi_1^0$ coupling is suppressed (however this suppression
may be somewhat offset by more phase space for the light $\chi _1^0$).
The dominant production
subprocesses $q\bar q'\to W^*\to\chi_1^+\chi_2^0$ and $q\bar q\to Z,\gamma\to
\chi_1^+\chi_1^-$ are then
determined mainly by the final state particle masses.
The masses of $\chi_1^0,\ \chi_2^0,\ \chi_1^+,\ h$ are illustrated in
Fig.~\ref{fig:mass1} versus $m_{1/2}$ for four typical choices of $m_0$ and the
sign of $\mu$. For the case $\mu > 0$ the
strongest phenomenological bound comes from the lightest
Higgs mass, which is known to be larger than about 60~GeV.  As has been
emphasized recently\cite{lp}, radiative
corrections are not known very accurately in this case as the
tree-level mass is very small, so we indicate this bound with a dashed line.
For the case $\mu < 0$ the strongest phenomenological bound comes from
the lightest chargino mass, which must be larger than about 45~GeV.
There are also weaker bounds coming from the naturalness condition
$|\mu| < 0.5$~TeV\cite{bbo801}; taken literally this condition would
exclude the higher $m_{1/2}$ ranges in Fig.~2, but since it is somewhat
subjective we do not apply it strictly.
The production cross sections for $\chi_1^\pm\chi_2^0,\ \chi_1^\pm\chi_1^0,\
\chi_1^+\chi_1^-$, obtained from the ISAJET program\cite{isasusy}, are  shown
in Fig.~\ref{fig:cs1}.

We address the situation where the superpotential contains a possible
lepton-number violating term $\lambda_{ijk} L_L^i L_L^j \bar E_R^k$ where $E_R$
is the
superfield containing the right-handed charged-lepton singlets and  $L_L$
contains the lepton doublets; the $i,j,k$ are
generation indices. These interactions break $R$-parity, since they involve an
odd number of supersymmetric particles. The Lagrangian generated by this
superpotential term has the form
\begin{equation}
  {\cal L} = \lambda_{ijk} \left[ \tilde\nu_L^i \bar e_R^k e_L^j + \tilde e_L^j
\bar e_R^k \nu_L^i + \left(\tilde e_R^k\right)^* \left(\bar\nu_L^i\right)^c
e_L^j - (i\leftrightarrow j) \right] + \rm h.c.
\end{equation}
in four-component Dirac notation.
A fundamental RPV interaction of this kind could mediate the decay of
any chargino or neutralino, as illustrated in Fig.~\ref{fig:feyns}(c) for
$\chi_i^0$. It is
natural to assume a hierarchy of
interactions such that RPC approximately
holds in decays of the heavier gauginos, and RPV
is manifest only in the otherwise forbidden
decay of the LSP $\chi_1^0$. Thus the next stage of our
analysis is to evaluate
the branching fractions for $\chi_1^+$ and $\chi_2^0$ RPC
decays, which can be found using the ISASUSY program\cite{isasusy}.
Figure~\ref{fig:br1} shows that the leptonic $\chi_1^+\to\chi_1^0\ell^+\nu$
fraction is typically 20--30\% (summing $\ell=e,\mu$), which sometimes goes
via $\chi_1^0 W^+, \tilde \ell^+\nu$ or $\ell^+\tilde\nu$ on-shell intermediate
states. Figure~\ref{fig:br2} shows that $\chi_2^0\to\chi_1^0\ell\ell$ is often
substantial; the branching fraction
sometimes depends on intermediate $\tilde\ell\ell$ states and
sometimes is suppressed due to competition with $\chi_2^0\to\chi_1^0 h$, which
always dominates when kinematically allowed.

For the RPV effects, we shall assume that the decay leptons do not
include $\tau$ and that the coupling is strong enough for $\chi^0_1$ to decay
near the production vertex, giving a lower limit to the RPV coupling
constant\cite{ross,dawson} $\lambda > (\gamma/20)(m_{\tilde
\ell}/100{\rm~GeV})^2 (1{\rm ~GeV}/m_{\rm LSP})^{2.5}$, where $\gamma$ is the
Lorentz boost factor of the LSP and $m_{\tilde \ell}$ is the mass of the
intermediate slepton in the $\chi_1^0\to \ell\ell'\nu$ decay. In our examples
below this condition implies $\lambda\agt10^{-4}$--$10^{-5}$.
We also require $\lambda\ll 1$ so that RPV terms do not significantly affect
the RGE and existing bounds on the couplings\cite{bhg} are respected.
One obtains the modified RGE at one-loop
\begin{eqnarray}
{{d\lambda _{ijk}}\over {dt}}&=&{{\lambda _{ijk}}\over {16\pi^2}}\left [
-{9\over 5}g_1^2-3g_2^2+(\delta_{j3}+2\delta_{k3})\lambda _\tau^2\right ]
+{1\over {16\pi^2}}\left [{\cal O}(\lambda _{lmn}^3)
\right ]\;, \label{lslashl}
\end{eqnarray}
where $\lambda _{ijk} $ are the various RPV couplings, and $\delta _{ij}$ is
the Kronecker delta.
The existing bounds\cite{bhg} require that the
$\lambda _{ijk}$ are small near
the electroweak scale, and then from Eq.~(\ref{lslashl}) it follows (for small
$\tan \beta$ considered here) that they
are small for all scales up to the GUT scale. Consequently the RPV
couplings $\lambda _{ijk}$ have a negligible effect on the
lepton Yukawa coupling running and on the running of the soft-supersymmetry
breaking parameters.
[The situation may be different
for the baryon-violating couplings, where the weaker electroweak scale bounds
together with the fixed-point character of the associated RGE can yield
large baryon-violating couplings near the GUT scale\cite{br}; these couplings
must be zero to avoid fast proton decay when lepton-violating couplings
are nonzero].  Finally, we have tacitly  assumed that the
$L_iL_j\bar E_k$ terms dominate over possible  $L_iQ_j\bar D_k$ RPV terms
\cite{dim,ross}; if the latter
are not negligible, the LSP can also decay into quarks and the multilepton
signals we present here become upper bounds.

  For multilepton acceptance we follow Ref.\cite{baer} and require rapidity
$|y(\ell)|<2.5$ for all leptons, transverse momentum $p_T(\ell)>15$~GeV for at
least one lepton, $p_T(\ell)>10$~GeV for the second lepton, and
$p_T(\ell)>8$~GeV for all other observed leptons. In the case of the spoiler
mode $\chi_2^0\to\chi_1^0h$, we assume the dominant $h\to jj$ dijet decay and
require all observed leptons to have separations $\Delta R =
\sqrt{(\Delta\eta)^2+(\Delta\phi)^2}>0.7$ from both of these jets. These
acceptance cuts give a semi-realistic estimate of how many leptons may indeed
be observable experimentally. We estimate acceptance factors by Monte Carlo
methods, using phase-space decay distributions with full kinematic constraints
from intermediate on-shell states.

The following four cases A--D illustrate interesting aspects of the RPC and RPV
multilepton signals in the four parameter regions (a)--(d) of
Figs.~\ref{fig:mass1}--\ref{fig:br2} above. All cases have $\tan\beta=1.5$.
We sum over lepton flavors $\ell=e, \mu$.

\noindent
{\bf Case A: {\boldmath$m_{1/2}={}$110~GeV, $m_0={}$200~GeV, $\mu <$~0.}}\\
 This gives $m(\chi_1^+)=65$~GeV, $m(\chi_2^0)=68$~GeV, $m(\chi_1^0)=34$~GeV,
$m(\tilde{g})=300$~GeV,
$m(h)=78$~GeV. With RPC there is a straightforward
$\chi_1^\pm\chi_2^0\to(\chi_1^0\ell\nu)(\chi_1^0\ell\ell)$ trilepton
signal (no spoiler mode, $\chi_2^0\not\rightarrow\chi_1^0 h$).
With RPV however, these final states give up to 7
leptons; also channels such as
$\chi_1^\pm\chi_2^0\to(\chi_1^0\ell\nu)(\chi_1^0\nu\nu)$ and
$\chi_1^\pm\chi_1^0\to\chi_1^0\chi _1^0\ell\nu$ give up to 5 leptons.
Our calculations give the following cross sections $\sigma_n$
for $\geq n$ leptons to pass acceptance cuts.
\[
\begin{array}{rlcll}
\mbox{RPC: \ } & \chi_1^\pm(\chi_2^0\to\chi_1^0\ell\ell) &\Rightarrow&
\sigma_3 &= 0.03 \rm~pb \;; \\
\mbox{RPV: \ } & \chi_1^\pm(\chi_2^0\to\chi_1^0\ell\ell) &\Rightarrow&
\sigma_3, \sigma_4, \sigma_5, \sigma_6, \sigma_7 &= 0.10, 0.08, 0.06, 0.025,
0.005 \rm~pb \;; \\
& \chi_1^\pm(\chi_2^0\to\chi_1^0\nu\nu) &\Rightarrow&
\sigma_3, \sigma_4, \sigma_5 &= 0.15, 0.08, 0.02 \rm~pb \;; \\
& \chi_1^\pm\chi_1^0 &\Rightarrow& \sigma_3, \sigma_4, \sigma_5 &= 0.19,
0.10, 0.03 \rm~pb \;; \\
& \chi_1^+\chi_1^- &\Rightarrow& \sigma_3, \sigma_4, \sigma_5, \sigma_6  &=
0.06, 0.05, 0.02, 0.005 \rm~pb \;.
\end{array}
\]
 This demonstrates that RPV can not only enhance the original RPC trilepton
channel but can produce quiet multileptons from other channels too.

\noindent
{\bf Case B: {\boldmath$m_{1/2}={}$110~GeV, $m_0={}$50~GeV, $\mu <$~0.}}\\
 This gives $m(\chi_1^+)=58$~GeV, $m(\chi_2^0)=66$~GeV, $m(\chi_1^0)=30$~GeV,
$m(\tilde{g})=300$~GeV,
$m(h)=72$~GeV, $m(\tilde\ell_R)=65$~GeV, so again there is no spoiler mode.
However, $\chi_2^0\to\chi_1^0\ell\ell$ decay proceeds almost entirely via
$\chi_2^0\to\tilde\ell\ell$ with $m(\tilde\ell)\approx m(\chi_2^0)$, for
which one final lepton is very soft and does not pass acceptance cuts,
so the RPC trilepton signal almost vanishes. We obtain
\[
\begin{array}{rlcll}
\mbox{RPC: \ } & \chi_1^\pm(\chi_2^0\to\chi_1^0\ell\ell) &\Rightarrow&
\sigma_3 &= 0.003 \rm~pb \;; \\
\mbox{RPV: \ } & \chi_1^\pm(\chi_2^0\to\chi_1^0\ell\ell) &\Rightarrow&
\sigma_3, \sigma_4, \sigma_5, \sigma_6 &= 0.38, 0.28, 0.14, 0.03 \rm~pb \;;
\\
& \chi_1^\pm(\chi_2^0\to\chi_1^0\nu\nu) &\Rightarrow&
\sigma_3, \sigma_4, \sigma_5 &= 0.28, 0.14, 0.03 \rm~pb \;; \\
& \chi_1^\pm\chi_1^0 &\Rightarrow& \sigma_3, \sigma_4, \sigma_5 &= 0.96, 0.43,
0.08 \rm~pb \;; \\
& \chi_1^+\chi_1^- &\Rightarrow& \sigma_3, \sigma_4, \sigma_5, \sigma_6  &=
0.24, 0.16, 0.07, 0.01 \rm~pb \;.
\end{array}
\]
In this example RPV rescues an otherwise suppressed trilepton signal and gives
up to 6 final leptons.  The biggest RPV signals here come from $\chi_1^{\pm}
\chi_1^0$ production, which exceeds $\chi_1^{\pm} \chi_2^0$ production in
this parameter region (see Fig.3).

\noindent
{\bf Case C: {\boldmath$m_{1/2} = {}$150~GeV, $m_0={}$200~GeV, $\mu >$~0.}}\\
This gives $m(\chi_1^+)=129$~GeV, $m(\chi_2^0)=129$~GeV, $m(\chi_1^0)=63$~GeV,
$m(\tilde{g})=410$~GeV,
$m(h)=61$~GeV, so the spoiler mode $\chi_2^0\to\chi_1^0h$ dominates and
suppresses the RPC trilepton signal. However, RPV gives multilepton plus $h\to
jj$ final states from this spoiler mode, and in principle gives quiet
multileptons from $\chi_1^\pm\chi_1^0$ and $\chi_1^+\chi_1^-$ production.
\[
\begin{array}{rlcll}
\mbox{RPC: \ } & \chi_1^\pm(\chi_2^0\to\chi_1^0\ell\ell) & \Rightarrow &
\sigma_3 &\simeq 0 \;; \\
\mbox{RPV: \ } & \chi_1^\pm(\chi_2^0\to\chi_1^0h) & \Rightarrow &
\sigma_3, \sigma_4, \sigma_5 &= 0.026, 0.017, 0.005 \rm~pb \;; \\
& \chi_1^\pm\chi_1^0 &\Rightarrow& \sigma_3, \sigma_4, \sigma_5 &= 0.001,
0.001, 0.000 \rm~pb \;; \\
& \chi_1^+\chi_1^- &\Rightarrow& \sigma_3, \sigma_4, \sigma_5, \sigma_6  &=
0.005, 0.005, 0.004, 0.002 \rm~pb \;.
\end{array}
\]
Here RPV rescues the normal spoiler mode.
The much smaller multilepton cross sections in this and the
following case D (both with $\mu>0$) are principally due to the much smaller
gaugino pair production cross sections; see Fig.3.

\noindent
{\bf Case D: {\boldmath$m_{1/2}={}$150~GeV, $m_0={}$50~GeV, $\mu >$~0.}}\\
 This case gives $m(\chi_1^\pm)=133$~GeV, $m(\chi_2^0)=133$~GeV,
$m(\chi_1^0)=65$~GeV, $m(\tilde{g})=410$~GeV,
$m(h)=59$~GeV, $m(\tilde\ell_R)=77$~GeV,
$m(\tilde\ell_L)=m(\tilde\nu)=119$~GeV. It is on the edge of being
excluded by the $m_h$ experimental bound, but we retain it to illustrate
a class of solutions.
Here the spoiler mode $\chi_2^0\to\chi_1^0h$ dominates but
$\chi_2^0\to\tilde\ell\ell\to\chi_1^0\ell\ell$ modes manage to compete thanks
to the on-shell intermediate state, so the conventional RPC trileptons are only
partially suppressed.  There is also an appreciable
$\chi_2^0\to\tilde\nu\nu\to\chi_1^0\nu\nu$ fraction . The branching
fraction $B(\chi_1^+\to \tilde\ell\nu (\ell\tilde\nu) \to \chi_1^0\ell\nu)$
= 0.65 is remarkably large too.  We obtain
\[
\begin{array}{rlcll}
\mbox{RPC: \ } & \chi_1^\pm(\chi_2^0\to\tilde\ell\ell\to\chi_1^0\ell\ell) &
\Rightarrow & \sigma_3 &= 0.003 \rm~pb \;; \\
\mbox{RPV: \ } & \chi_1^\pm(\chi_2^0\to\chi_1^0 h) &
\Rightarrow & \sigma_3, \sigma_4, \sigma_5 &= 0.027, 0.018, 0.006 \rm~pb \;;\\
& \chi_1^\pm(\chi_2^0\to\tilde\ell\ell\to\chi_1^0\ell\ell) &
\Rightarrow & \sigma_3, \sigma_4, \sigma_5, \sigma_6, \sigma_7 &= 0.005, 0.005,
0.005, 0.004, 0.002 \rm~pb \;; \\
& \chi_1^\pm(\chi_2^0\to\tilde\nu\nu\to\chi_1^0\nu\nu) & \Rightarrow &
\sigma_3, \sigma_4, \sigma_5 &= 0.009, 0.008, 0.004 \rm~pb \;;\\
& \chi_1^\pm\chi_1^0 & \Rightarrow & \sigma_3, \sigma_4, \sigma_5 &= 0.002,
0.002, 0.001 \rm~pb \;;\\
& \chi_1^+\chi_1^- &\Rightarrow& \sigma_3, \sigma_4, \sigma_5, \sigma_6  &=
0.026, 0.025, 0.021, 0.010 \rm~pb \;.
\end{array}
\]
In this example, RPV rescues the dominant spoiler contributions and also gives
quiet multileptons in other channels.

These illustrative RPV signals total $\sigma_3=$ 0.5, 1.9, 0.03,
0.07~pb, in cases A, B, C, D respectively, for events with $n \geq 3$
leptons.  The recent CDF trilepton search\cite{cdf}, based on 11.1~pb$^{-1}$
luminosity, found just two candidate events; however, their
acceptance criteria appear to be more restrictive than ours,
and our analysis omits detection-efficiency factors.  It is possible
therefore that RPV in Case B is already excluded by data,
but it still provides an instructive example.  Future high-statistics
data will put stronger constraints on parameters in the RPV scenario.

To summarize, we have investigated purely leptonic RPV signatures in
chargino-neutralino pair production at the Tevatron, using typical
SUSY-GUT parameter sets taken from Ref.\cite{bbo801} in the low-$\tan\beta$
fixed-point region.  These examples illustrate how RPV
decays $\chi_1^0\to\ell\ell'\nu$  convert the quiet trilepton production
of the RPC scenario into multilepton signals and generate new signals
in other final states. In particular they show that

\noindent
i) the trilepton rate itself may be greatly enhanced;

\noindent
ii) signals with 4 or 5 leptons are typically comparable with the
unsuppressed RPC trilepton rate;

\noindent
iii) when RPC trileptons are suppressed by the spoiler mode
$\chi_2^0\to\chi_1^0h$ or by $\chi_2^0\to\tilde\ell\ell$ with $m(\tilde\ell)
\simeq m(\chi_2^0)$, RPV multilepton signals can still be substantial,
though some contain $h\to jj$ dijets;

\noindent
iv) substantial quiet RPV multilepton signals can also come from
$\chi_1^\pm(\chi_2^0\to\chi_1^0\nu\nu),\ \chi_1^\pm\chi_1^0$ and
$\chi_1^+\chi_1^-$ channels, that give no RPC trileptons.

\newpage
\acknowledgements

We thank H.~Baer, J.~Beacom and E.~Mirkes for discussions on technical aspects
of the calculations. This research was supported
in part by the University of Wisconsin Research Committee with funds granted by
the Wisconsin Alumni Research Foundation, in part by the U.S.~Department of
Energy under Contract No.~DE-AC02-76ER00881, and in part by the Texas National
Laboratory Research Commission under Grant Nos.~RGFY93-221 and FCFY9302.
MSB was supported in part by an SSC Fellowship. PO was supported in
part by an NSF Graduate Fellowship.

\newpage
\section*{Figures}

\begin{enumerate}

\item{Representative diagrams for (a)~the pair production of charginos and
neutralinos, (b)~their RPC decay, and (c)~the RPV decay of the lightest
neutralino. \label{fig:feyns}}

\item{Masses of the particles in $\chi_1^+$, $\chi_2^0$ decay cascades,
obtained from Ref.\cite{bbo} for representative GUT-scale inputs.
Note that $\tilde e_L = \tilde e_2$ and $\tilde e_R = \tilde e_1$ because the
mixing angle is negligible.
\label{fig:mass1}}

\item{Cross sections for chargino and neutralino pair production at $\sqrt s =
1.8$~TeV, for the SUSY-GUT parameters of Fig.~\ref{fig:mass1}.
\label{fig:cs1}}

\item{Decay branching fractions of the lightest chargino
$\chi_1^+$. \label{fig:br1}}

\item{Decay branching fractions of the second-lightest neutralino
$\chi_2^0$. \label{fig:br2}}

\end{enumerate}

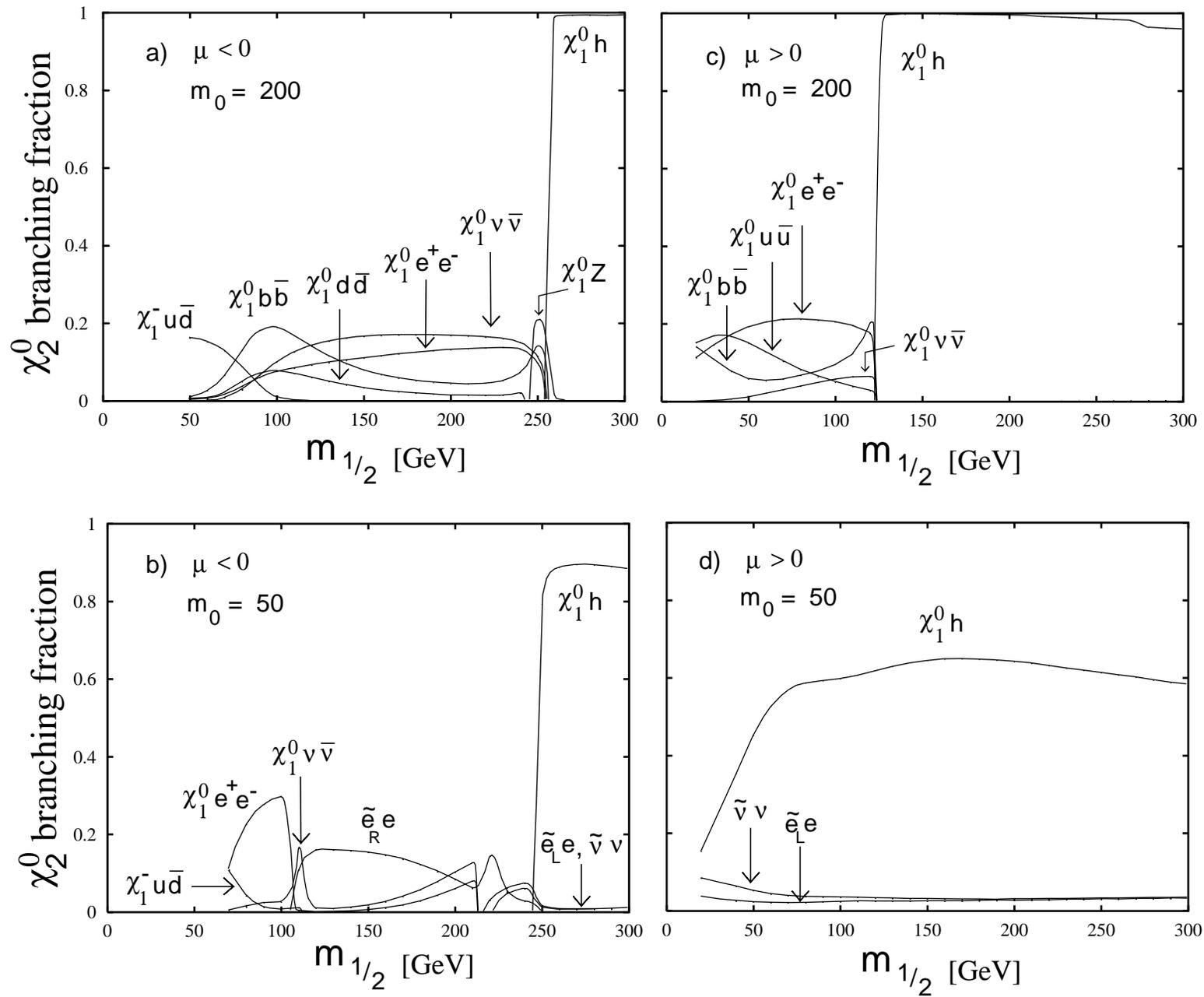
\end{document}